\documentclass[letter]{aa}

\usepackage{graphicx}
\usepackage{txfonts}
\usepackage{natbib}
\bibpunct{(}{)}{;}{a}{}{,}
\begin{document}

\title{Serendipitous VLBI detection of rapid, large-amplitude, \\ intraday
  variability in QSO 1156+295}

  \author{T.~Savolainen\inst{1}
           \and
          Y.~Y.~Kovalev\inst{1,2}}

  \offprints{T.~Savolainen}

  \institute{Max-Planck-Institut f\"ur Radioastronomie, Auf dem
             H\"ugel 69, D-53121, Bonn, Germany\\
             \email{tsavolainen@mpifr-bonn.mpg.de,
             ykovalev@mpifr-bonn.mpg.de} \and
             Astro Space Center of Lebedev Physical Institute,
             Profsoyuznaya 84/32, 117997 Moscow, Russia} 

\date{Received $<$date$>$; accepted $<$date$>$}

\abstract{}{We report a serendipitous detection of rapid, large amplitude flux
  density variations in the highly core-dominated, flat-spectrum radio quasar
  1156+295 during an observing session at the Very Long Baseline Array
  (VLBA).}{The source was observed as a part of the MOJAVE survey programme
  with the VLBA at 15\,GHz on February 5, 2007. Large amplitude variability in
  the correlated flux density, unexplainable in terms of the source structure,
  was first discovered while processing the data, and later confirmed by
  calibrating the antenna gains using 24 other sources observed in the
  experiment.}{The source shows variations in the correlated flux density as
  high as 40\% on a timescale of only 2.7 hours. This places 1156+295 between
  the classical IDV sources and the so-called intra-hour variables. The
  observed variability timescale and the modulation index of 13\% are
  consistent with interstellar scintillation by a nearby, highly turbulent
  scattering screen. The large modulation index at 15\,GHz implies a
  scattering measure that is atypically high for a high galactic latitude
  source such as 1156+295.}{}


  \titlerunning{Large-amplitude, intraday variability in 1156+295}
 
  \maketitle

%

\section{Introduction}

Intraday variability (IDV) of compact, radio-loud AGN at cm-wavelengths was
first discovered in the mid-1980s \citep{wit86,hee87} and both
source-intrinsic and extrinsic mechanisms for IDV have been vigorously studied
\citep[for a review, see e.g.][]{wag95}. Detection of time delays in the
variability pattern arrival times between widely separated telescopes, as well
as observed annual modulation of the variability timescale, have shown
conclusively that interstellar scintillation (ISS) is the cause of intra-hour
flux density variations observed in the three most extreme IDV sources
\object{PKS\,0405-385} \citep{jau00}, \object{J1819+3845} \citep{den02,den03},
and \object{PKS\,1257-326} \citep{big03,big06}. Evidence for the ISS origin of
IDV was found also in the cases of \object{0917+624}
\citep{ric01,jau01,fur02}, \object{PKS\,1519-273} \citep{jau03}, and
\object{J1128+5925} \citep{gab07}. This provided the possibility to study both
small-scale spatial fluctuations in the interstellar medium and the structure
of compact radio sources on the microarcsecond scale. Unfortunately, the
extreme IDV sources, which are the most suitable for these studies, are
rare. By ``extreme'', we mean sources showing variability on timescales of a
few hours or less and with an rms amplitude of modulation of over 10\%. In
this Letter, we report the serendipitous discovery of very large amplitude IDV
in quasar \object{1156+295} with a timescale of variations shorter than
3\,h. This discovery is also unusual because it was achieved by a VLBI
experiment.

\begin{figure}
 \centering
   \resizebox{0.75\hsize}{!}{\includegraphics{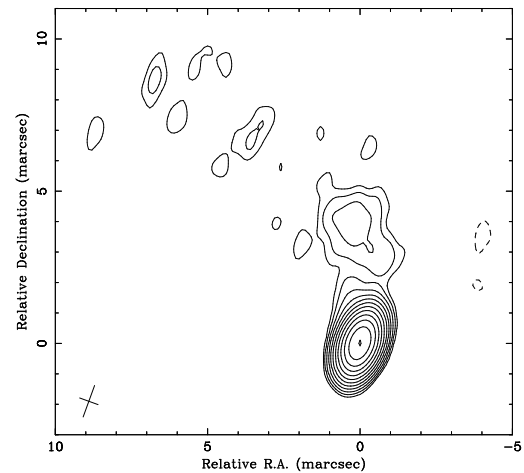}}
   \caption{Naturally weighted 15\,GHz VLBA image of \object{1156+295}
     observed on February 5, 2007. The map peak brightness is
     1.46\,Jy\,beam$^{-1}$. The contours begin at 0.7\,mJy\,beam$^{-1}$ and
     increase in steps of 2. Since the source flux density varied
     significantly during the observation, an amplitude self-calibration had
     to be used early in the imaging process. This may affect the image
     quality.}
   \label{map}
\end{figure}

\begin{figure*}
 \centering
   \resizebox{0.9\hsize}{!}{\includegraphics{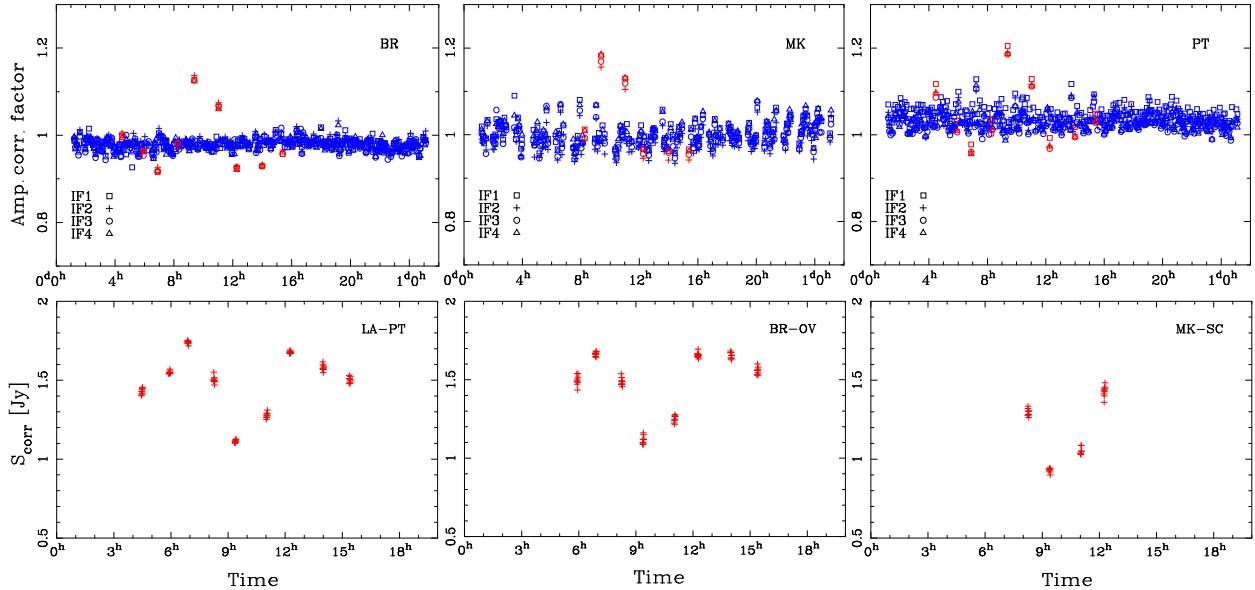}}
   \caption{{\it Top:} Amplitude self-calibration solutions of three example
     antennas (Brewster, Mauna Kea, and Pie Town) for the VLBA experiment on
     February 5, 2007. There is one solution per scan, and IFs are kept
     separate. Low elevation scans ($<15$\degr) are excluded from the
     figure. Red symbols correspond to scans of \object{1156+295} and blue
     symbols correspond to all other sources. The obvious discrepancy between
     the solutions for \object{1156+295} and the other sources is present at
     all ten antennas.  {\it Bottom:} Examples of the correlated flux density
     curves of \object{1156+295} at three baselines after calibration of
     antenna gains by amplitude correction factors interpolated from the
     nearby scans of other sources. Time is in UT.}
   \label{calib}
\end{figure*}

\object{1156+295} (\object{4C\,+29.45}) is an optically violently variable
quasar at $z=0.729$ \citep{bur68}, which has Galactic coordinates
$l=199.4$\degr, $b=+78.4$\degr. The source is strongly variable throughout the
electromagnetic spectrum from radio to $\gamma$-rays.  At radio frequencies,
\object{1156+295} shows significant long-term variability on a timescale of
months \citep[e.g.][]{kov02}. \citet{wil83} reported a range of at least 5
magnitudes in the optical brightness, and in gamma-rays \citet{sre96} detected
an order of magnitude variability using the EGRET instrument onboard the
Compton Gamma Ray Observatory.  During high radio brightness states,
\object{1156+295} appears to be highly core-dominated in VLBI images (95\% at
some epochs \citep{kov05}; see also Fig.~\ref{map}). Superluminal motion at
speeds ranging from $3.5 h^{-1} c$ to $14.1 h^{-1} c$ were reported by
e.g. \citet{pin97}, \citet{jor01}, \citet{hon04}, and \citet{kel04}.

In the optical bands, \object{1156+295} shows rapid variations on timescales
of between less than 30 minutes and 3 days \citep{wil83,rai98}. The first IDV
detection of \object{1156+295} at radio frequencies was reported by
\citet{lov03}, who observed the source with the VLA at 5\,GHz in January 2002.
Their study measured rms flux density variations of 167\,mJy during 3 nights
of observations and a variability timescale of $\sim24$\,h. The relative
amplitude of variations observed by \citet{lov03} was far smaller than
reported here.

Throughout this paper, we use the following cosmological parameters: $H_0$ =
73 km s$^{-1}$ Mpc$^{-1}$, $\Omega_M$ = 0.27, and $\Omega _\Lambda$ = 0.73.
The angular scaling conversion for $z=0.729$ is 7.07\,pc mas$^{-1}$.

\section{Observations and analysis}

Quasar \object{1156+295} and 24 other compact radio sources were observed with
the NRAO's Very Long Baseline Array (VLBA) at 15\,GHz as a part of the MOJAVE
project \citep{lis05} on February 5, 2007\footnote{We note that
  \object{1156+295} was observed several times in the MOJAVE survey and its
  IDV behaviour during the rest of the MOJAVE epochs will be analysed in
  Kuchibhotla et al. (in prep.).}. The entire observing session had a duration
of 24 hours, which included 9 scans of \object{1156+295}, each lasting 4.7
minutes. The signal was recorded in dual-polarisation mode: 4 IFs (each with
8\,MHz bandwidth) per circular polarisation and 1-bit sampling. The data from
the experiment were calibrated following standard procedures of VLBI data
reduction \citep[see e.g.][]{lis05}. After {\it a priori} amplitude
calibration and fringe fitting, the correlated flux density of
\object{1156+295} showed significant, correlated, temporal variability from
scan to scan on \emph{every baseline}: between 7h and 12h UT, it first dropped
by about 0.6\,Jy and then increased again. The dip can be seen on every
baseline, which excludes the source structure as the cause of the
variations. The system temperature measurements used in the {\it a priori}
amplitude calibration do not show any jumps that could explain the dip.

To ascertain whether the variability is genuine or due to amplitude
calibration problems, we first imaged and self-calibrated the $(u,v)$ data of
\object{1156+295} in a normal manner and then compared the resulting antenna
gain correction factors with those derived from the self-calibration of the
other 24 sources observed in the same session. Amplitude self-calibration with
a 5-minute solution interval was indeed able to remove the variability of
\object{1156+295}, but it produced gain amplitude correction factors that were
-- \emph{for every antenna} -- significantly offset from the gain corrections
determined from the 24 other sources in the experiment (see examples in
Fig.~\ref{calib}).  This can be explained if the variability is genuine but
does not alter the shape of the source's visibility function significantly
during the experiment, e.g. if the variability occurs in one dominant compact
component. Since the antenna gain is not source-dependent, calibration errors
cannot account for the above-described behaviour.

Although improbable, there is a possibility that the observed variations could
be due to a source-specific correlator problem. In such a case, however, one
would expect to see discrepant solutions from the fringe fitting. We carefully
checked the fringe-fit solutions and could not find any anomalies. Therefore,
an instrumental effect that originates in the correlator seems highly
unlikely. Taken together, we conclude from the above arguments that the
variations in the correlated flux density of \object{1156+295} are due to
genuine intraday variability observed in the source.

\begin{figure}
 \centering
   \resizebox{0.94\hsize}{!}{\includegraphics{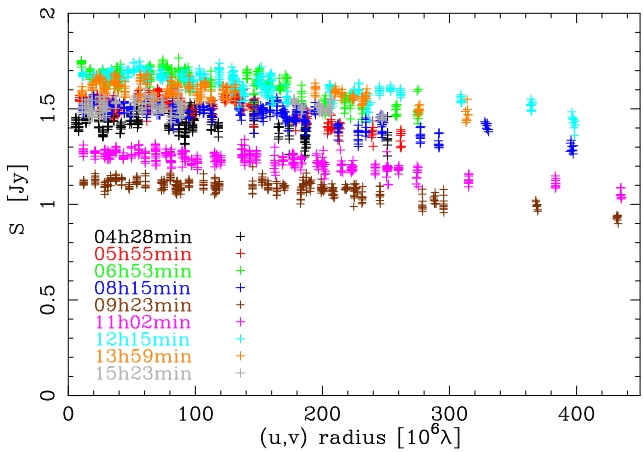}}
   \caption{Calibrated correlated Stokes I flux density of \object{1156+295}
     (in Jy) as a function of $(u,v)$ radius (in M$\lambda$) for each
     individual scan of the VLBI experiment on February 5, 2007. The data has
     been averaged over the IFs and for 30\,s in time. The different scans are
     shown in different colours with the scan start time in UT indicated in
     the legend.}
   \label{uvrad}
\end{figure}

To study rigorously the observed IDV event in \object{1156+295}, the residual
errors in the antenna gains, which remain after the {\it a priori} amplitude
calibration, need to be corrected. Since a large number of sources were
observed during the VLBA session with their scans interleaved, it was possible
to estimate the antenna gain correction factors for the scans of
\object{1156+295} by linearly interpolating the amplitude self-calibration
solutions from the scans of the other sources that were observed within one
hour of each scan of \object{1156+295}. Before interpolation, the scans at
elevations below $15\degr$ were removed. The bottom panels in Fig.~\ref{calib}
show examples of correlated flux density curves on three baselines that were
calibrated in this way. The variations in the correlated flux density are very
similar at short (LA-PT), intermediate (BR-OV), and long (MK-SC) baselines.
Again, this is expected, if the variability occurs in a single dominant
compact component. In Fig.~\ref{uvrad}, we plot the calibrated correlated flux
density of \object{1156+295} as a function of $(u,v)$ radius for each
individual scan. The figure shows that the shape of the visibility function,
$\mathcal{V}(r_{u,v})$, changes little from scan to scan, although the
correlated flux density changes by 0.6\,Jy. To analyse possible small changes
in the shape of $\mathcal{V}(r_{u,v})$, we constructed the mean visibility
function of the observation, $\langle \mathcal{V}(r_{u,v}) \rangle $, by
averaging the normalised visibilities of all individual scans. The correlated
flux density of each scan was first binned into $50$\,M$\lambda$ wide bins in
projected $(u,v)$ spacing and then normalised to 1.0 at
$r_{u,v}=25$\,M$\lambda$.  We subtracted $\langle \mathcal{V}(r_{u,v})
\rangle$ from the binned $\mathcal{V}(r_{u,v})$ of each individual scan and
inspected the residuals. Some variability was found: $\mathcal{V}(r_{u,v})$ at
$200 \lesssim r_{u,v} \lesssim 300$\,M$\lambda$ increased by 8\% during the
observation, which is more than the $\approx 3$\% variability expected from
the time variable response of the interferometer to the faint jet. However,
the variability turned out to be solely due to the baselines to St. Croix, and
therefore probably was due to calibration inaccuracies of the
antenna. Consequently, we are unable to confirm conclusively the detection of
time variability in $\mathcal{V}(r_{u,v})$. An upper limit to variability is
8\%.

\begin{figure}
 \centering
   \resizebox{0.94\hsize}{!}{\includegraphics{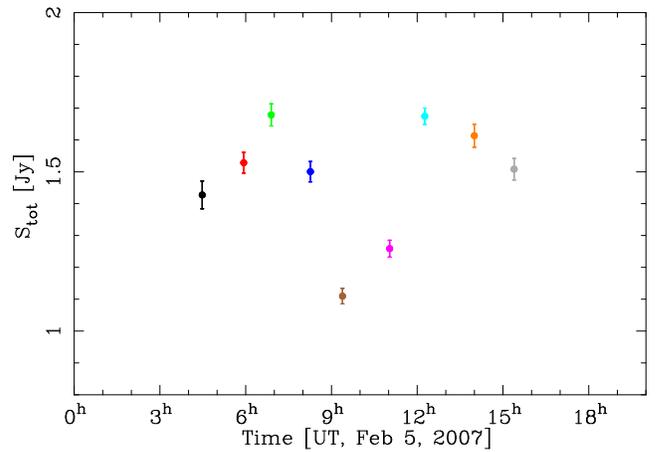}}
   \caption{Integrated flux density curve of \object{1156+295}, obtained by
     averaging the correlated flux density at projected baselines shorter than
     100\,M$\lambda$. The error bars represent the standard deviation of the
     correlated flux density (at baselines $<100$M$\lambda$) in each
     scan. Colour coding of the scans is the same as in
       Fig.~\ref{uvrad}.}
   \label{lc}
\end{figure}

We constructed an integrated VLBA flux density curve for \object{1156+295} by
averaging the correlated flux density at projected baselines between
$6-100$\,M$\lambda$ (Fig.~\ref{lc}). Since $\mathcal{V}(r_{u,v})$ was
essentially flat in this range (Fig.~\ref{uvrad}), the averaging provided a
good estimate of the integrated emission coming from angular scales
$\lesssim30$\,mas. Averaging also reduces the errors due to inaccurate
amplitude calibration, because these are antenna-specific.

The two important quantities in the analysis of IDV are the modulation index
and the characteristic timescale of variability.  We consider the average of
the peak-to-trough and trough-to-peak times of the large dip to be the
timescale of variability, which is therefore $t_\mathrm{var}=2.7\pm0.5$\,h. We
note that our $t_\mathrm{var}$ is approximately a factor of 1.7 times longer
than the timescale on which the autocorrelation function of intensity
fluctuations reaches a fraction $1/e$ of its maximum value \citep{jau01}.  The
modulation index $m$, defined to be the standard deviation of the source flux
density divided by the mean source flux density, is $13\pm3$\% for the flux
density curve in Fig.~\ref{lc}. The uncertainties in $t_\mathrm{var}$ and $m$
were estimated following the analysis in \citet{den03}. The modulation index
is much higher than the values typically observed for IDV sources at
frequencies above 5\,GHz \citep{ked01,kra03}. For \object{1156+295},
\cite{lov03} reported $m=5.8$\% at 5\,GHz in 2002.

\section{Discussion}

If we assumed a source-intrinsic origin of the observed variability in
\object{1156+295}, it would imply, by light travel time arguments, a
brightness temperature of $\gtrsim 2 \times 10^{19}$\,K, which is far in
excess of the inverse Compton (IC) limit of $10^{12}$\,K \citep{kel69}. A
Doppler factor higher than $\sim 270$ would be then required to avoid the IC
catastrophe. This would cause severe problems with the energy requirements in
the source \citep{beg94}. Extremely fast jets are also unlikely in the light
of VLBI monitoring surveys, which indicate that the maximum jet Lorentz factor
$\Gamma$ is $30-40$ \citep{coh07}. The standard model of incoherent
synchrotron radiation from relativistic electrons in the jet would therefore
have significant difficulties in explaining the observed variability, if it
were intrinsic. For this reason, we concentrate on two possible extrinsic
causes: interstellar scintillation and an extreme scattering event.

\begin{figure}
 \centering
   \resizebox{1.00\hsize}{!}{\includegraphics{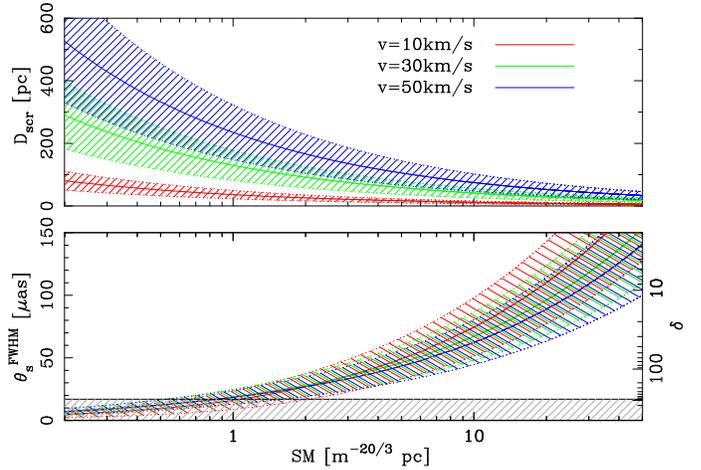}}
   \caption{Parameters of the possible scattering screen. {\it Top:} The
     distance to the screen as a function of the scattering measure {\it
       SM}. Different screen velocities are shown in different colours. {\it
       Bottom:} The scintillating source size as a function of {\it SM}. The
     right edge of the panel gives the Doppler factor if
     $\theta_\mathrm{s}^{FWHM}$ is equal to equipartition size. The grey
     shaded area corresponds to $\theta_\mathrm{s}^{FWHM} < 17$\,$\mu$as. In
     both panels, the hatched region around the solid lines gives 1$\sigma$
     error for the plotted quantities. The +1$\sigma$ lines in the
     $\theta_\mathrm{s}^{FWHM}$ plot roughly correspond to -1$\sigma$ lines in
     the $D_\mathrm{scr}$ plot and vice versa.}
   \label{model}
\end{figure}

Propagation effects in the ionised medium of the Milky Way will cause a
sufficiently compact source to scintillate \citep[for review of ISS, see
e.g.][]{ric90,goo97}. Since the observed $m$ in \object{1156+295} is clearly
below 100\%, it is reasonable to assume that diffractive scintillation is
fully quenched and to consider only refractive ISS. The observed
$t_\mathrm{var}$ and $m$ constrain the properties of a possible scattering
screen and the size of the scintillating source. Because of the high $m$ in
our case, it appears likely that our observing frequency is close to the
critical frequency $\nu_\mathrm{s}$ between strong and weak scattering
regimes. \citet{goo97} provided a formula for $m$ that can be used to
interpolate solutions close to $\nu_\mathrm{s}$, where the asymptotic
solutions for strong and weak regimes break down. Combining his Eqs.  10, 12,
18, 19, and 20, and assuming a Gaussian brightness profile for the source, we
calculated the screen distance, $D_\mathrm{scr}$, and the FWHM size of the
scintillating source $\theta_\mathrm{s}^{FWHM}$ as a function of scattering
measure \mbox{{\it SM} $=\int_0^d C_N^2 dx$}, where $C_N^2$ is the strength of
the electron-density fluctuations\footnote{Note that $\theta_\mathrm{s}$ in
  \citet{goo97} is approximately $\theta_\mathrm{s}^{FWHM} / 2.35$.}
(Fig.~\ref{model}). The results were parameterised by the screen velocity. We
estimated the uncertainties in $D_\mathrm{scr}$ and $\theta_\mathrm{s}^{FWHM}$
by Monte Carlo methods and the hatched regions in Fig.~\ref{model} show the
1$\sigma$ error for these quantities. In addition, the results given by the
interpolation formula of \citet{goo97} were confirmed by numerically
integrating the intensity covariance function for refractive scintillation
\citep[see Appendix A in][]{ric06}. In the bottom panel of Fig.~\ref{model},
we indicate an approximate lower limit to the source size,
$\theta_\mathrm{s}^{FWHM} \ge 17$\,$\mu$as, given by the IC catastrophe limit
and by assuming an upper limit of 50 for the Doppler factor
\citep{lah99,coh07}. As can be seen from Fig.~\ref{model}, if the variability
is due to ISS, we have either a rather nearby screen ($D_\mathrm{scr} \lesssim
300$\,pc) with {\it SM} $\gtrsim 0.5$\,m$^{-20/3}$\,pc \emph{or}
$\theta_\mathrm{s}^{FWHM}$ so small that it would require Doppler factor
exceeding 50.

We can estimate $\theta_\mathrm{s}^{FWHM}$ independently by assuming that
there is approximate equipartition between the energy densities of the
magnetic field and the radiating electrons \citep{sco77}. We assume the
synchrotron peak frequency to be 15\,GHz and the corresponding peak flux
density to be 1.5\,Jy. This is based on the assumption that the scintillating
component is the core of a \citet{bla79} type jet and corresponds to the
$\tau=1$ surface at our observing frequency. The resulting equipartition size
is $\theta_s^{FWHM} \sim 270 \cdot \delta^{-7/17}$\,$\mu$as, where $\delta$ is
the Doppler factor. If we again take $\delta \le 50$, this results in
$\theta_\mathrm{s}^{FWHM} \gtrsim 55$\,$\mu$as, {\it SM}\,$\gtrsim
4$\,m$^{-20/3}$\,pc, and $D_\mathrm{scr} \lesssim 100$\,pc for screen
velocities between 10 and 50\,km\,s$^{-1}$. Since the Galactic electron
distribution model by \citet{cor02} predicts a scattering measure of only
0.1\,m$^{-20/3}$\,pc for the line-of-sight towards \object{1156+295}, our
results indicate a nearby, localised region of highly turbulent ionised gas in
that direction.

The frequency-dependence of $m$ is slightly surprising: one would not expect
$m$ to increase from 5.8\% at 5\,GHz \citep{lov03} to 13\% at 15\,GHz, unless
the scattering is strong \emph{and} the source is smaller than the scattering
angle. In that case the scintillating component would contain only about
$30-50$\% of the total flux density of the core, because otherwise the
observed $m$ would be significantly higher. Another possibility is that
\object{1156+295} was in a more compact stage during our observation than it
was 5 years earlier.  Simultaneous multi-frequency measurements of $m$ are
required to clarify this.

It is also possible that the observed variations are not due to standard ISS,
but instead correspond to an isolated extreme scattering event
\citep[ESE;][]{fie87}. Unfortunately, our short, single-frequency observation
does not provide us with sufficient information to distinguish between these
two cases. However, the shape of the flux density curve resembles an ESE with
two maxima symmetrically surrounding a deep minimum. Although ``classical''
ESEs have durations of between weeks and months, \citet{cim02} reported an
event in \object{0954+658} with a duration of less than 2 days. The extremely
short timescale of our event would, in the case of an ESE, imply a cloud size
$<0.01$\,AU.

\begin{acknowledgements}
  We thank Lars Fuhrmann, David Jauncey, Richard Porcas, Eduardo Ros and the
  anonymous referee for their careful reading of the manuscript and their
  constructive comments. We also thank Thomas Krichbaum, Matthew Lister and
  Sarma Kuchibhotla for useful discussions. Both authors are research fellows
  of the Alexander von Humboldt Foundation. TS was also partially supported by
  the Max-Planck-Gesellschaft and by the Academy of Finland grant 120516. The
  National Radio Astronomy Observatory is a facility of the National Science
  Foundation operated under cooperative agreement by Associated Universities,
  Inc.
\end{acknowledgements}

\bibliographystyle{aa}
\bibliography{idv}

\begin{thebibliography}{42}
\expandafter\ifx\csname natexlab\endcsname\relax\def\natexlab#1{#1}\fi

\bibitem[{{Begelman} {et~al.}(1994){Begelman}, {Rees}, \& {Sikora}}]{beg94}
{Begelman}, M.~C., {Rees}, M.~J., \& {Sikora}, M. 1994, \apjl, 429, L57

\bibitem[{{Bignall} {et~al.}(2003){Bignall}, {Jauncey}, {Lovell}, {Tzioumis},
  {Kedziora-Chudczer}, {Macquart}, {Tingay}, {Rayner}, \& {Clay}}]{big03}
{Bignall}, H.~E., {Jauncey}, D.~L., {Lovell}, J.~E.~J., {et~al.} 2003, \apj,
  585, 653

\bibitem[{{Bignall} {et~al.}(2006){Bignall}, {Macquart}, {Jauncey}, {Lovell},
  {Tzioumis}, \& {Kedziora-Chudczer}}]{big06}
{Bignall}, H.~E., {Macquart}, J.-P., {Jauncey}, D.~L., {et~al.} 2006, \apj,
  652, 1050

\bibitem[{{Blandford} \& {K\"onigl}(1979)}]{bla79}
{Blandford}, R.~D. \& {K\"onigl}, A. 1979, \apj, 232, 34

\bibitem[{{Burbidge}(1968)}]{bur68}
{Burbidge}, E.~M. 1968, \apjl, 154, L109

\bibitem[{{Cim{\`o}} {et~al.}(2002){Cim{\`o}}, {Beckert}, {Krichbaum},
  {Fuhrmann}, {Kraus}, {Witzel}, \& {Zensus}}]{cim02}
{Cim{\`o}}, G., {Beckert}, T., {Krichbaum}, T.~P., {et~al.} 2002, PASA, 19, 10

\bibitem[{{Cohen} {et~al.}(2007){Cohen}, {Lister}, {Homan}, {Kadler},
  {Kellermann}, {Kovalev}, \& {Vermeulen}}]{coh07}
{Cohen}, M.~H., {Lister}, M.~L., {Homan}, D.~C., {et~al.} 2007, \apj, 658, 232

\bibitem[{{Cordes} \& {Lazio}(2002)}]{cor02}
{Cordes}, J.~M. \& {Lazio}, T.~J.~W. 2002, arXiv:astro-ph/0207156

\bibitem[{{Dennett-Thorpe} \& {de Bruyn}(2002)}]{den02}
{Dennett-Thorpe}, J. \& {de Bruyn}, A.~G. 2002, \nat, 415, 57

\bibitem[{{Dennett-Thorpe} \& {de Bruyn}(2003)}]{den03}
{Dennett-Thorpe}, J. \& {de Bruyn}, A.~G. 2003, \aap, 404, 113

\bibitem[{{Fiedler} {et~al.}(1987){Fiedler}, {Dennison}, {Johnston}, \&
  {Hewish}}]{fie87}
{Fiedler}, R.~L., {Dennison}, B., {Johnston}, K.~J., {et~al.} 1987,
  \nat, 326, 675

\bibitem[{{Fuhrmann} {et~al.}(2002){Fuhrmann}, {Krichbaum}, {Cim{\`o}},
  {Beckert}, {Kraus}, {Witzel}, {Zensus}, {QIAN}, \& {Rickett}}]{fur02}
{Fuhrmann}, L., {Krichbaum}, T.~P., {Cim{\`o}}, G., {et~al.} 2002, PASA, 19, 64

\bibitem[{{Gab{\'a}nyi} {et~al.}(2007){Gab{\'a}nyi}, {Marchili}, {Krichbaum},
  {Britzen}, {Fuhrmann}, {Witzel}, {Zensus}, {M{\"u}ller}, {Liu}, {Song},
  {Han}, \& {Sun}}]{gab07}
{Gab{\'a}nyi}, K.~{\'E}., {Marchili}, N., {Krichbaum}, T.~P., {et~al.} 2007,
  \aap, 470, 83

\bibitem[{{Goodman}(1997)}]{goo97}
{Goodman}, J. 1997, New Astronomy, 2, 449

\bibitem[{{Heeschen} {et~al.}(1987){Heeschen}, {Krichbaum}, {Schalinski}, \&
  {Witzel}}]{hee87}
{Heeschen}, D.~S., {Krichbaum}, T., {Schalinski}, C.~J., {et~al.} 1987,
  \aj, 94, 1493

\bibitem[{{Hong} {et~al.}(2004){Hong}, {Jiang}, {Gurvits}, {Garrett},
  {Garrington}, {Schilizzi}, {Nan}, {Hirabayashi}, {Wang}, \&
  {Nicolson}}]{hon04}
{Hong}, X.~Y., {Jiang}, D.~R., {Gurvits}, L.~I., {et~al.} 2004, \aap, 417, 887

\bibitem[{{Jauncey} {et~al.}(2003){Jauncey}, {Johnston}, {Bignall}, {Lovell},
  {Kedziora-Chudczer}, {Tzioumis}, \& {Macquart}}]{jau03}
{Jauncey}, D.~L., {Johnston}, H.~M., {Bignall}, H.~E., {et~al.} 2003, \apss,
  288, 63

\bibitem[{{Jauncey} {et~al.}(2000){Jauncey}, {Kedziora-Chudczer}, {Lovell},
  {Nicolson}, {Perley}, {Reynolds}, {Tzioumis}, \& {Wieringa}}]{jau00}
{Jauncey}, D.~L., {Kedziora-Chudczer}, L.~L., {Lovell}, J.~E.~J., {et~al.}
  2000, in Astrophysical Phenomena Revealed by Space VLBI, ed.
  H.~{Hirabayashi}, P.~G. {Edwards}, \& D.~W. {Murphy}, 147--150

\bibitem[{{Jauncey} \& {Macquart}(2001)}]{jau01}
{Jauncey}, D.~L. \& {Macquart}, J.-P. 2001, \aap, 370, L9

\bibitem[{{Jorstad} {et~al.}(2001){Jorstad}, {Marscher}, {Mattox}, {Wehrle},
  {Bloom}, \& {Yurchenko}}]{jor01}
{Jorstad}, S.~G., {Marscher}, A.~P., {Mattox}, J.~R., {et~al.} 2001, \apjs,
  134, 181


\bibitem[{{Kedziora-Chudczer} {et~al.}(2001){Kedziora-Chudczer}, {Jauncey},
  {Wieringa}, {Tzioumis}, \& {Reynolds}}]{ked01}
{Kedziora-Chudczer}, L.~L., {Jauncey}, D.~L., {et~al.} 2001, \mnras, 325, 1411

\bibitem[{{Kellermann} {et~al.}(2004){Kellermann}, {Lister}, {Homan},
  {Vermeulen}, {Cohen}, {Ros}, {Kadler}, {Zensus}, \& {Kovalev}}]{kel04}
{Kellermann}, K.~I., {Lister}, M.~L., {Homan}, D.~C., {et~al.} 2004, \apj, 609,
  539

\bibitem[{{Kellermann} \& {Pauliny-Toth}(1969)}]{kel69}
{Kellermann}, K.~I. \& {Pauliny-Toth}, I.~I.~K. 1969, \apjl, 155, L71

\bibitem[{{Kovalev} {et~al.}(2005){Kovalev}, {Kellermann}, {Lister}, {Homan},
  {Vermeulen}, {Cohen}, {Ros}, {Kadler}, {Lobanov}, {Zensus}, {Kardashev},
  {Gurvits}, {Aller}, \& {Aller}}]{kov05}
{Kovalev}, Y.~Y., {Kellermann}, K.~I., {Lister}, M.~L., {et~al.} 2005, \aj,
  130, 2473

\bibitem[{{Kovalev} {et~al.}(2002){Kovalev}, {Kovalev}, {Nizhelsky}, \&
    {Bogdantsov}}]{kov02} {Kovalev}, Y.~Y., {Kovalev}, Y.~A., {Nizhelsky},
  N.~A. {et~al.} 2002, PASA, 19, 83

\bibitem[{{Kraus} {et~al.}(2003){Kraus}, {Krichbaum}, {Wegner}, {Witzel},
  {Cim{\`o}}, {Quirrenbach}, {Britzen}, {Fuhrmann}, {Lobanov}, {Naundorf},
  {Otterbein}, {Peng}, {Risse}, {Ros}, \& {Zensus}}]{kra03}
{Kraus}, A., {Krichbaum}, T.~P., {Wegner}, R., {et~al.} 2003, \aap, 401, 161

\bibitem[{{L{\"a}hteenm{\"a}ki} \& {Valtaoja}(1999)}]{lah99}
{L{\"a}hteenm{\"a}ki}, A. \& {Valtaoja}, E. 1999, \apj, 521, 493

\bibitem[{{Lister} \& {Homan}(2005)}]{lis05}
{Lister}, M.~L. \& {Homan}, D.~C. 2005, \aj, 130, 1389

\bibitem[{{Lovell} {et~al.}(2003){Lovell}, {Jauncey}, {Bignall},
  {Kedziora-Chudczer}, {Macquart}, {Rickett}, \& {Tzioumis}}]{lov03}
{Lovell}, J.~E.~J., {Jauncey}, D.~L., {Bignall}, H.~E., {et~al.} 2003, \aj,
  126, 1699

\bibitem[{{Piner} \& {Kingham}(1997)}]{pin97}
{Piner}, B.~G. \& {Kingham}, K.~A. 1997, \apjl, 485, L61

\bibitem[{{Raiteri} {et~al.}(1998){Raiteri}, {Ghisellini}, {Villata}, {de
  Francesco}, {Lanteri}, {Chiaberge}, {Peila}, \& {Antico}}]{rai98}
{Raiteri}, C.~M., {Ghisellini}, G., {Villata}, M., {et~al.} 1998, \aaps, 127,
  445

\bibitem[{{Rickett}(1990)}]{ric90}
{Rickett}, B.~J. 1990, \araa, 28, 561

\bibitem[{{Rickett} {et~al.}(2001){Rickett}, {Witzel}, {Kraus}, {Krichbaum}, \&
  {Qian}}]{ric01}
{Rickett}, B.~J., {Witzel}, A., {Kraus}, A., {et~al.} 2001, \apjl, 550, L11

\bibitem[{{Rickett} {et~al.}(2006){Rickett}, {Lazio}, \& {Ghico}}]{ric06}
{Rickett}, B.~J., {Lazio}, T.~J.~W., \& {Ghico}, F.~D. 2006, \apjs, 165, 439

\bibitem[{{Scott} \& {Readhead}(1977)}]{sco77}
{Scott}, M.~A. \& {Readhead}, A.~C.~S. 1977, \mnras, 180, 539

\bibitem[{{Sreekumar} {et~al.}(1996){Sreekumar}, {Bertsch}, {Dingus},
  {Esposito}, {Fichtel}, {Fierro}, {Hartman}, {Hunter}, {Kanbach}, {Kniffen},
  {Lin}, {Mayer-Hasselwander}, {Mattox}, {Michelson}, {von Montigny},
  {Mukherjee}, {Nolan}, {Schneid}, {Thompson}, \& {Willis}}]{sre96}
{Sreekumar}, P., {Bertsch}, D.~L., {Dingus}, B.~L., {et~al.} 1996, \apj, 464,
  628


\bibitem[{{Wagner} \& {Witzel}(1995)}]{wag95}
{Wagner}, S.~J. \& {Witzel}, A. 1995, \araa, 33, 163

\bibitem[{{Wills} {et~al.}(1983){Wills}, {Pollock}, {Aller}, {Aller},
  {Balonek}, {Barvainis}, {Binzel}, {Chaffee}, {Dent}, {Douglas}, {Fanti},
  {Garrett}, {Gregorini}, {Henry}, {Hill}, {Howard}, {Jeske}, {Kepler},
  {Leacock}, {Mantovani}, {O'Dea}, {Padrielli}, {Perley}, {Pica}, {Puschell},
  {Sanduleak}, {Shields}, {Smith}, {Thuan}, {Wade}, {Wasilewski}, {Webb},
  {Wills}, \& {Wisniewski}}]{wil83}
{Wills}, B.~J., {Pollock}, J.~T., {Aller}, H.~D., {et~al.} 1983, \apj, 274, 62

\bibitem[{{Witzel} {et~al.}(1986){Witzel}, {Heeschen}, {Schalinski}, \&
  {Krichbaum}}]{wit86}
{Witzel}, A., {Heeschen}, D.~S., {Schalinski}, C., \& {Krichbaum}, T. 1986,
  Mitteilungen der Astronomischen Gesellschaft Hamburg, 65, 239

\end{thebibliography}

\end{document}